\documentclass[twocolumn,showpacs,preprintnumbers,amsmath,amssymb,superscriptaddress]{revtex4}
\usepackage[bookmarks, pdftitle={}, pdfauthor={Nair}, colorlinks=true, linkcolor=black, citecolor=blue, urlcolor=black]{hyperref}
\usepackage{amssymb}
\usepackage{amsmath}
\usepackage{graphicx}% Include figure files
\usepackage{dcolumn}% Align table columns on decimal point
\usepackage{bm}% bold math
\usepackage{epstopdf}
\usepackage{times}
\begin{document}
\preprint{}
\title
{Spin-Lattice Coupling and Frustrated Magnetism in Fe-doped Hexagonal LuMnO$_3$}
\author{Harikrishnan S. Nair}
\email{h.nair.kris@gmail.com, hsnair@uj.ac.za}
\affiliation{Highly Correlated Matter Research Group, Physics Department, P. O. Box 524, University of Johannesburg, Auckland Park 2006, South Africa}
\author{Zhendong Fu}
\affiliation{J\"{u}lich Center for Neutron Sciences JCNS, Outstation at MLZ, Forschungszentrum J\"{u}lich GmbH, Lichtenberg Stra{\ss}e 1, D-85747 Garching, M\"{u}nchen, Germany}
\author{C. M. N. Kumar}
\affiliation{J\"{u}lich Centre for Neutron Science JCNS, Outstation at SNS, Oak Ridge National Laboratory, Oak Ridge, Tennessee 37831, United States}
\affiliation{Chemical and Engineering Materials Division, Oak Ridge National Laboratory, Oak Ridge, Tennessee 37831, United States}
\author{V. Y. Pomjakushin}
\affiliation{Laboratory for Neutron Scattering, Paul Scherrer Institute, CH-5232 Villigen, Switzerland }
\author{Yinguo Xiao}
\affiliation{J\"{u}lich Center for Neutron Sciences JCNS and Peter Gr\"{u}nberg Institute PGI, JARA-FIT, Forschungszentrum J\"{u}lich GmbH, 52425 J\"{u}lich, Germany}
\author{Tapan Chatterji}
\affiliation{Institut Laue-Langevin, BP 156, 38042 Grenoble Cedex 9, France}
\author{Andr\'{e} M. Strydom}
\affiliation{Highly Correlated Matter Research Group, Physics Department, P. O. Box 524, University of Johannesburg, Auckland Park 2006, South Africa}
\affiliation{Max Planck Institute for Chemical Physics of Solids, N\"{o}thnitzerstra{\ss}e 40, 01187 Dresden, Germany}
%
%\author{Ramesh Kumar K.}
%\affiliation{Highly Correlated Matter Research Group, Physics Department, P. O. Box 524, University of Johannesburg, Auckland Park 2006, South Africa}
%
\date{\today}
\begin{abstract}
Strong spin-lattice coupling and prominent frustration effects observed in the 50$\%$ Fe-doped frustrated hexagonal ($h$)LuMnO$_3$ are reported. A N\'{e}el transition at $T_{\mathrm N} \approx$ 112~K and a possible spin re-orientation transition at $T_{\mathrm {SR}} \approx$ 55~K are observed in the magnetization data. From neutron powder diffraction data, the nuclear structure at and below 300~K was refined in polar $P6_3cm$ space group. While the magnetic structure of LuMnO$_3$ belongs to the $\Gamma_4$ ($P6'_3c'm$) representation, that of LuFe$_{0.5}$Mn$_{0.5}$O$_3$ belongs to $\Gamma_1$ ($P6_3cm$) which is supported by the strong intensity for the $\mathbf{(100)}$ reflection and also judging by the presence of spin-lattice coupling. The refined atomic positions for Lu and Mn/Fe indicate significant atomic displacements at $T_{\mathrm N}$ and $T_{\mathrm {SR}}$ which confirms strong spin-lattice coupling. Our results complement the discovery of room temperature multiferroicity in thin films of $h$LuFeO$_3$ and would give impetus to study LuFe$_{1-x}$Mn$_x$O$_3$ systems as potential multiferroics where electric polarization is linked to giant atomic displacements.
\end{abstract}
\maketitle
\indent 
Hexagonal manganites $(h)R$MnO$_3$ ($R$ = rare earth) are fascinating systems in the class of multifunctional oxides which present multiferroicity\cite{lottermoser_nature_430_2004magnetic,leenature_451_805_2008,vanakennaturematerials_3_164_2004}, dielectric and magnetic anomalies,\cite{katsufuji_prb_64_2001dielectric} field-induced re-entrant phases,\cite{lorenzprl_92_087204_2004} metamagnetic steps in magnetization\cite{nair_prb_83_2011effect} and recently, even found related to cosmology\cite{griffin_prx_2_041022_2012scaling}. The primary interest in hexagonal manganites arises from the potential to realize multiferroics since it was found that below the ferroelectric transition temperature, $T_{\mathrm{FE}} \approx$ 1000~K, they develop electric polarization due to structural distortions and giant atomic displacements\cite{vanakennaturematerials_3_164_2004,leenature_451_805_2008}. $hR$MnO$_3$ systems often have a low N\'{e}el temperature $T_{\mathrm{N}}$ (often $\leq$ 100~K) compared to the $T_{\mathrm{FE}}$. Strong coupling between lattice, magnetic and electric degrees of freedom is generally observed in hexagonal manganites despite the large separation in temperature between $T_{\mathrm{FE}}$ and $T_N$\cite{fiebignature_419_818_2002}. The magnetic structure of $hR$MnO$_3$ presents significant complexity in the form of magnetic frustration where the Mn moments in $ab$-plane form 120$^{\circ}$ triangular lattice\cite{brownjpcm_18_10085_2006neutron}. This 2D edge-sharing triangular network is geometrically frustrated with antiferromagnetic nearest-neighbor exchange interaction and gives rise to diffuse magnetic scattering intensity close to $T_{\mathrm N}$\cite{parkprb_68_104426_2003magnetic}. 
\\
\indent
There exist significantly many reports on the magnetic structure of hexagonal manganites in different representations due to the homometric pairs of irreducible represenations that yield the same neutron diffraction pattern if the $x$(Mn) is the ideal value of ($\frac{1}{3}$). One of the most-studied hexagonal manganite, YMnO$_3$ has been reported in several symmetries by different authors -- $\Gamma_3$\cite{fiebig_prl_2000determination}, $\Gamma_1$\cite{fabreges_prl_2009spin} or $\Gamma_5$, $\Gamma_6$\cite{brownjpcm_18_10085_2006neutron}. Earlier investigations suggested $\Gamma_1$ or $\Gamma_3$ as the possible magnetic structure for YMnO$_3$\cite{munoz_prb_62_2000magnetic}. However, neutron polarimetry studies have confirmed the magnetic structure as $\Gamma_5$ or $\Gamma_6$\cite{brownjpcm_18_10085_2006neutron}. On the other hand, $h$LuMnO$_3$ which shares most of the properties of $h$YMnO$_3$ has $\Gamma_4$ ($P6'_3c'm$) symmetry\cite{solovyev_prb_86_2012magnetic,parkprb_68_104426_2003magnetic}. In both $h$YMnO$_3$ and $h$LuMnO$_3$, strong spin-lattice coupling triggered by giant atomic displacements of Mn ion at $T_{\mathrm N}$ plays an important role in inducing ferroelectric distortion leading subsequently to multiferroic property.
\\
\indent
Compared to $h$LuMnO$_3$, the ferrite $h$LuFeO$_3$ is reported to present improved magnetic properties\cite{das_nature_5_2014bulk}. In a recent study on $h$LuFeO$_3$, strong exchange coupling leading to high magnetic transition temperatures and spin re-orientation transitions closely connected with structural distortions have been identified\cite{wang_prb_90_2014structural}. Thin films of $h$LuFeO$_3$ have shown evidence for a room temperature multiferroic\cite{wang_prl_110_2013room} and has been identified as a strong candidate for linear magnetoelectric coupling and control of the ferromagnetic moment directly by an electric field\cite{disseler_arxiv2014magnetic}. However, bulk LuFeO$_3$ normally crystallizes in orthorhombic $Pbnm$ symmetry thereby precluding the possibility of ferroelectricity. Hence it is desirable to prepare solid solutions of $hR$MnO$_3$ and $hR$FeO$_3$ in order to combine the features of giant atomic displacements that lead to ferroelectric polarization and the magnetic transitions at near-300~K. In the present work, the solid solution LuFe$_{0.5}$Mn$_{0.5}$O$_3$ has been synthesized and is studied using magnetization and neutron powder diffraction. The results reveal strong spin-lattice coupling in this material and suggest its potential to be a multiferroic.
\\
%
% 
% EXPERIMENTAL
\indent
The polycrystalline samples used in the present study were prepared by following a solid state reaction method\cite{katsufuji_prb_64_2001dielectric}. High purity Lu$_2$O$_3$, Fe$_2$O$_3$ and MnO (4$N$, Sigma Aldrich) were used as the precursors. The synthesized powder sample was first characterized using laboratory x rays. Formation of a single hexagonal phase without impurities was confirmed in this way. Magnetic measurements were performed using a Magnetic Property Measurement System, Quantum Design Inc. Neutron powder diffraction experiments were carried out at the SINQ spallation source of Paul Scherrer Institute (Switzerland) using the high-resolution diffractometer for thermal neutrons HRPT\cite{hrpt} with the wavelength $\lambda = 1.886$~\AA\ in high intensity mode. About 8~g of powder sample was used to obtain the neutron powder patterns which were recorded at 11 temperature points between 2~K and 300~K. The nuclear and magnetic structure refinements were performed using Rietveld method\cite{rietveld} employing FULLPROF code\cite{carvajal}. Magnetic structure refinement using representation analysis was performed using the software SARA$h$\cite{wills_sarah}.
\\
% #######################################################################################################
\begin{figure}[!t]
\centering
\includegraphics[scale=0.35]{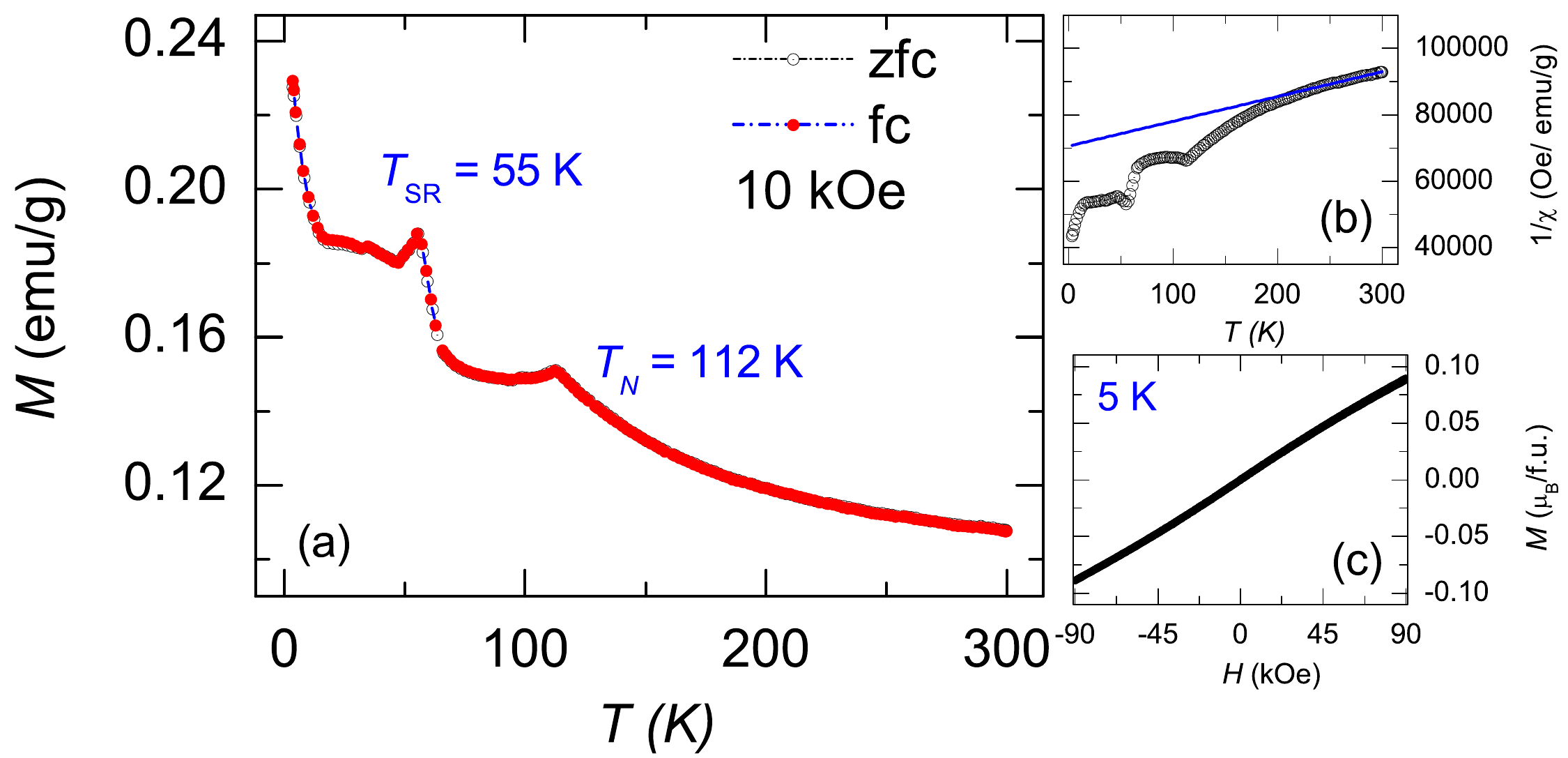}
\caption{(colour online) (a) Magnetization curves of LuFe$_{0.5}$Mn$_{0.5}$O$_3$ in zero field-cooled (zfc) and field-cooled (fc) modes at an applied field of 10~kOe. Two magnetic transitions occur at $T_{\mathrm N}\approx$ 112~K and $T_{\mathrm {SR}}\approx$ 55~K. (b) Presents the inverse susceptibility suggesting presence of spin fluctuations at $T \geq T_{\mathrm N}$. The solid line is Curie-Weiss fit. (c) Shows field-scans of magnetization at 5~K.}
\label{fig_mt}
\end{figure}
% #######################################################################################################
%
%
% RESULTS: MAGNETIZATION
\indent
Magnetization measurements in zero field-cooled (zfc) and field-cooled (fc) protocols recorded for LuFe$_{0.5}$Mn$_{0.5}$O$_3$ are presented in Fig~\ref{fig_mt} (a) for 10~kOe. Two magnetic phase transitions are identified at $T_{\mathrm N}\approx$ 112~K and $T_{\mathrm {SR}} \approx$ 55~K. At $T_N$, the N\'{e}el transition in the 120$^{\circ}$ triangular lattice takes place. In the case of $h$LuMnO$_3$ a lower value of $T_{\mathrm N}\approx$ 88~K was observed. The transition at $T_{\mathrm {SR}}$ in LuFe$_{0.5}$Mn$_{0.5}$O$_3$ could be a spin re-orientation transition similar to the one found in $h$LuFeO$_3$ at 130~K\cite{wang_prb_90_2014structural}. The $T_{\mathrm {SR}}$ in $h$LuFeO$_3$ is closely related to the interlayer exchange coupling and the atomic displacements due to the $K_1$ phonon mode\cite{wang_prb_90_2014structural}. The frustration effects in LuFe$_{0.5}$Mn$_{0.5}$O$_3$ are clearly seen in the inverse magnetic susceptibility 1$/\chi (T)$ which shows deviation from linear trend even for $T \gg T_{\mathrm N}$, as seen from the Curie-Weiss fit in (b). From the Curie-Weiss fit, an effective paramagnetic moment value, $\mu_{\mathrm{eff}}$ = 5.41(4)~$\mu_{\mathrm B}$ and Curie-Weiss temperature, $\Theta_{\mathrm {CW}}$ = -946~K are estimated. The effective paramagnetic moment calculated assuming spin-only contributions is $\mu_{\mathrm{calc}}$ = 5.4~$\mu_B$. As an estimate of frustration, the ratio $f$ = $|\Theta_{\mathrm {CW}}|$/$T_{\mathrm N}\approx$ 8.5 is calculated. This value of $f$ signals significant frustration effects and is comparable to the frustration indices of other hexagonal systems collected in Table~\ref{tab0}. A field-scan performed in zero field-cooled protocol at 5~K is presented in Fig~\ref{fig_mt} (c) where no hysteresis is observed. There is no indication of ferromagnetic contribution to magnetic susceptibility. The maximum magnetic moment obtained at 5~K under 90~kOe ($\approx$ 0.1~$\mu_\mathrm{B}/\mathrm{f.u.}$) is significantly reduced in magnitude compared to the value for the ferromagnetic alignment of Mn$^{3+}$ and Fe$^{3+}$ moments; 5.4~$\mu_B/{\mathrm {f.u.}}$. The antiferromagnetic arrangement of moments in the basal plane and the resulting strong frustration effects leads to this reduction in observed magnetic moment. In addition to the purely geometrical frustration effects, low dimensionality of the hexagonal plane brought about by the competition between the intra-plane and the inter-plane exchange interactions, also plays a role in the separation between $T_N$ and $\Theta_{CW}$.  
\\
%
% #######################################################################################################
\begin{table*}
\centering
\caption{\label{tab0} The Curie-Weiss temperature, $\Theta_{CW}$, the N\'{e}el temperature, $T_N$ and the frustration parameter, $f$ of some of the highly frustrated hexagonal manganites. The values for LuFe$_{0.5}$Mn$_{0.5}$O$_3$ closely compare with those for other related hexagonal systems.}
\begin{tabular}{lcccll} \hline\hline
Compound & $\Theta_{CW}$ (K)   &    $T_N$ (K)  &   $f$     &  Ref. \\ \hline\hline
YMnO$_3$  &  -545   & 75  & 7.8  &  \cite{ramirez2001handbook}&  \\
LuMnO$_3$ & -740 & $\approx$ 90 & $\approx$ 8 &  \cite{park_prb_82_2010doping}& \\
(Y,Lu)MnO$_3$ & -600 to -800 & $\approx$ 70 to 90 & $\approx$ 8 &  \cite{park_prb_82_2010doping}& \\
Lu(Fe,Mn)O$_3$ & -946 &  112 &  8.5 &[This Work] & \\ \hline\hline
\end{tabular}
\end{table*}
% #######################################################################################################
%
%
% RESULTS: NEUTRONS
%
%
% #######################################################################################################
\begin{table*}
\centering
\caption{\label{tab1} The refined lattice parameters and atomic positions of LuFe$_{0.5}$Mn$_{0.5}$O$_3$ at 300~K and 2~K. The refinement was carried out using the nuclear space group $P6_3cm$ ($\#$185). The atomic positions are: Lu1 $2a$ (0,0,$z$);  Lu2 $4b$ ($\frac{1}{3}$,$\frac{2}{3}$,$z$); Fe/Mn $6c$ ($x$,0,$z$); O1 $6c$ ($x$,0,$z$): O2 $6c$ ($x$,0,$z$); O3 $2a$ (0,0,$z$); O4 $4b$ ($\frac{1}{3}$,$\frac{2}{3}$,0). The isotropic thermal parameters were fixed at the values obtained from \cite{munoz_prb_62_2000magnetic}.}
\begin{tabular}{lllllllll}	\hline\hline
                            & & &  &     300~K   & & &  &       2~K         \\ \hline
$a (\AA)$            & & &  &   6.0094(5)                & & &  &     5.9962(9)                  \\
$c (\AA)$            & & &  &   11.5440(24)               & & &  &    11.5415(31)                     \\
Lu1: $z$                     & & &  &  0.213(46) & & &  &  0.090(51)            \\
Lu2: $z$                   & & &  &  0.172(33)  & & &  &  0.049(12) \\
Fe/Mn: $x$     & & &  &  0.348(117)   & & &  &  0.335(28)  \\
$z$               & & &  &  -0.057(57)   & & &  &  -0.179(22)  \\
O1: $x$        & & &  &  0.308(23)         & & &  &    0.308(18)  \\
$z$              & & &  &  0.072(32)      & & &  &   0.072(12)   \\
O2: $x$      & & &  &  0.640(28)  & & &  &   0.640(25)  \\
$z$              & & &  &   0.242(34)  & & &  &  0.242(23)  \\
O3: $z$            & & &  &  0.373(37) & & &  &   0.373(22)      \\
O4: $z$          & & &  &  -0.078(45)    & & &  & -0.078(24)       \\  \hline\hline
\end{tabular}
\end{table*}
% #######################################################################################################
%
% #######################################################################################################
\begin{figure}[!t]
\centering
\includegraphics[scale=0.45]{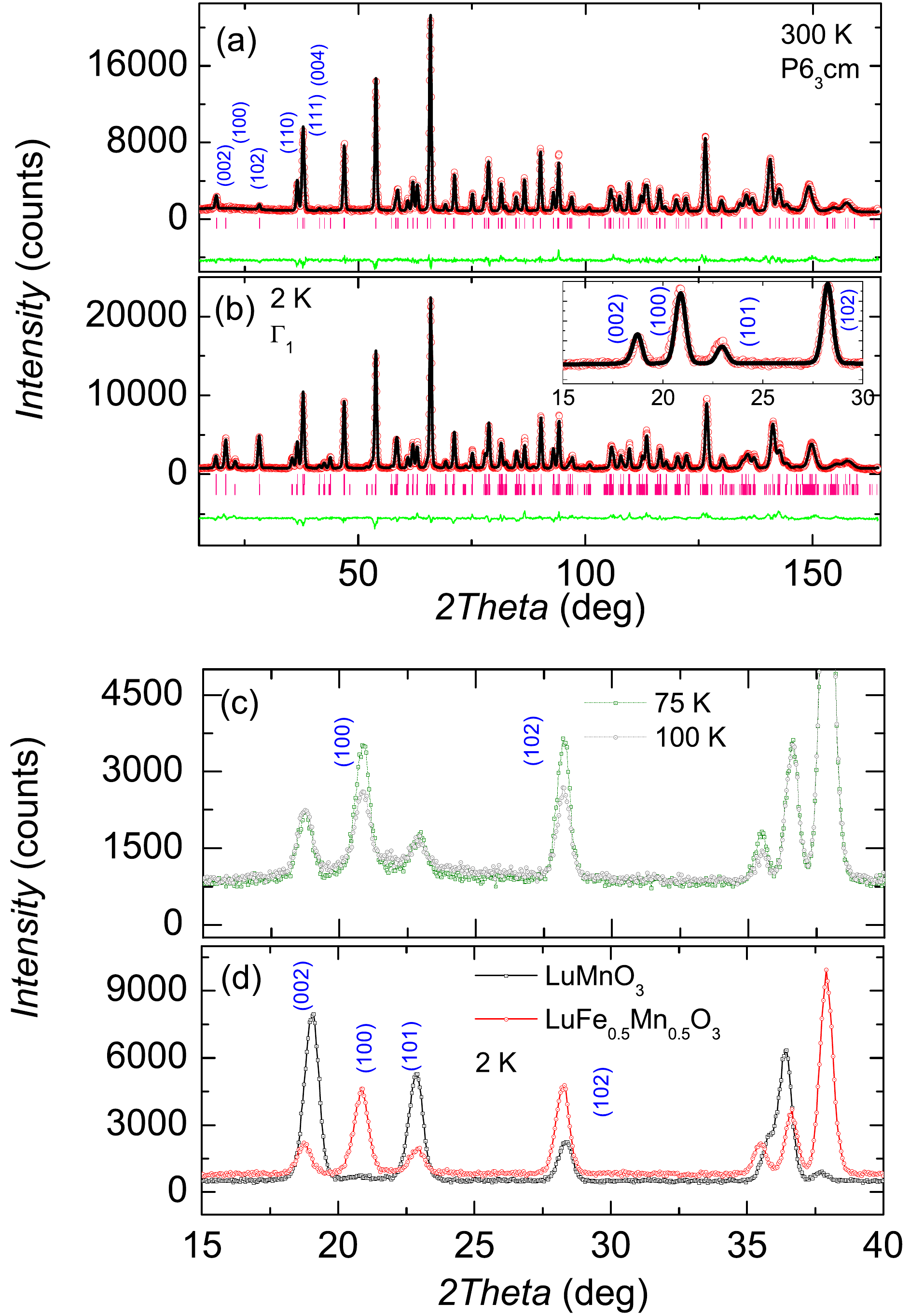}
\caption{\label{fig_nd}(colour online) The neutron powder diffraction data on LuFe$_{0.5}$Mn$_{0.5}$O$_3$ along with Rietveld refinement for (a) 300~K and (b) 2~K. The nuclear structure is refined in $P6_3cm$ space group. In (b), the magnetic structure is modeled after $\Gamma_1$. The calculated pattern is plotted as a black line and the difference plot as a green line. The Bragg positions are marked as pink vertical bars. In (b), the bottom row of ticks represent magnetic peaks. The inset of (b) magnifies the low-angle reflections, especially the $\mathbf{(100)}$, which is typically strong for $\Gamma_1$. (c) Shows the plot comparing the patterns at 75~K and 100~K which highlight an enhancement of intensity for the reflections $\mathbf{(100)}$ and $\mathbf{(102)}$. (d) Shows a comparison of the 2~K data of $h$LuMnO$_3$ and LuFe$_{0.5}$Mn$_{0.5}$O$_3$ to contrast $\Gamma_1$ ($\Gamma_3$) and $\Gamma_2$ ($\Gamma_4$).}
\end{figure}
% #######################################################################################################
\indent
The experimentally obtained neutron powder diffraction pattern for LuFe$_{0.5}$Mn$_{0.5}$O$_3$ at 300~K and 2~K are presented in Fig~\ref{fig_nd} (a) and (b) respectively. The calculated patterns refined using the Rietveld method are also shown. It is known that the hexagonal manganites undergo a phase transition from centrosymmetric $P6_3/mmc$ to ferroelectric $P6_3cm$ below $T_{\mathrm{FE}} \approx$ 1000~K\cite{vanakennaturematerials_3_164_2004}. It is found that the nuclear structure of LuFe$_{0.5}$Mn$_{0.5}$O$_3$ remains $P6_3cm$ in the temperature range 300 - 2~K. The refined lattice parameters and the atomic coordinates for 300~K and 2~K are presented in Table~\ref{tab1}. The refined Mn position at 10~K is $x$ = 0.334. For a perfect 2D triangular network, the ideal value is $x$ = $\frac{1}{3}$ and it is reported to be 0.340 for $h$YMnO$_3$ and 0.331 for $h$LuMnO$_3$\cite{park_prb_82_2010doping}. The displacements of the Mn atom as suggested by the $x$ position have strong correlation with the magnetic structure. The Mn-Mn interactions between adjacent Mn planes are due to superexchange mechanism via the apical oxygens of MnO$_5$ bipyramids. When $x$ = ($\frac{1}{3}$), all exchange paths are equivalent. However, when $x\neq$ ($\frac{1}{3}$), two different paths with two different exchnage interactions, $J_{z1}$ and $J_{z2}$, are formed. Thus $x$ = ($\frac{1}{3}$) is a critical threshold value and determines the stability of magnetic structure below $T_N$\cite{fabreges_prl_2009spin,das_nature_5_2014bulk}.
\\
\indent
Below $T_{\mathrm{N}} \approx$ 112~K, a purely magnetic reflection is observed at $\mathbf{(101)}$ at 2$\Theta \approx$ 23$^\circ$ ($d_{hkl}$ = 4.72~$\AA$) and enhancement of nuclear intensity at $\mathbf{(102)}$ at 2$\Theta \approx$ 28$^\circ$ ($d_{hkl}$ = 3.85~$\AA$; see Fig~\ref{fig_nd} (b), inset) suggesting antiferromagnetic ordering in the triangular lattice with Mn/Fe moments aligned 120$^{\circ}$ to each other. The magnetic structure of LuFe$_{0.5}$Mn$_{0.5}$O$_3$ below $T_{\mathrm N}$ was solved by assuming $\mathbf{k}$ = $\mathbf{(0,0,0)}$ propagation vector for the nuclear space group $P6_3cm$. Representation analysis for magnetic structure then allows six possible solutions: $\Gamma_1$ ($P6_3cm$), $\Gamma_2$ ($P6_3c'm'$), $\Gamma_3$ ($P6'_3cm'$), $\Gamma_4$ ($P6'_3c'm$), $\Gamma_5$ ($P6_3$) and $\Gamma_6$ ($P6_3'$)\cite{brownjpcm_18_10085_2006neutron,solovyev_prb_86_2012magnetic}. The magnetic structure of $h$LuMnO$_3$ belongs to the representation $\Gamma_4$ ($P6'_3c'm$)\cite{parkprb_68_104426_2003magnetic,solovyev_prb_86_2012magnetic}. However, non-zero intensity for the $\mathbf{(100)}$ reflection which is stronger compared to  the $\mathbf{(101)}$ could suggest that the $\Gamma_3$ ($P6'_3cm'$) or $\Gamma_1$ ($P6_3cm$) models are the correct one for the Fe-doped compound. The 2~K data with the Rietveld refinement fits assuming $\Gamma_1$ model is presented in Fig~\ref{fig_nd} (b). The indices of low-angle nuclear and magnetic reflections are marked in the figure. A part of the diffraction data is magnified in the inset of (b) to show the magnetic peaks and the fits. A comparison of the diffraction patterns obtained at 75~K and 100~K which lie between the $T_{\mathrm N}$ and the $T_{\mathrm {SR}}$ is presented in the panel (c). The reflections at $\mathbf{(100)}$ and $\mathbf{(102)}$ are seen to undergo an enhancement in intensity with the reduction in temperature. This could be an indication of the progressive change in the magnetic structure below the $T_\mathrm{N}$. The magnetic refinement above the $T_\mathrm{SR}$ was performed using the $\Gamma_1$ ($P6_3cm$) and the $\Gamma_3$ ($P6'_3cm'$) representations and was found that they both gave equivalent description of the observed data. Analysis was also carried out using the representation $\Gamma_2$ ($P6_3c'm'$) however, it could not reproduce the experimental magnetic intensities faithfully. Within the present set of measurements, it is not possible to clearly confirm a spin re-orientation at $T_\mathrm{SR}$ in this compound. In Fig~\ref{fig_nd} (d) a comparison of the 2~K data of $h$LuMnO$_3$ and LuFe$_{0.5}$Mn$_{0.5}$O$_3$ is presented. Note that the $\mathbf{(100)}$ peak is absent in the $h$LuMnO$_3$ data.
\\
Trimerization instability,\cite{leenature_451_805_2008,park_prb_82_2010doping} single ion anisotropy and Dzyaloshinksii-Moriya interactions are anticipated for the $P6_3cm$ structure and could play a role in determining the magnetic ground state. The details of how all these factors lead to a complex magnetic structure is yet to be understood. It was noted in the work by Solovyev {\em et al.,}\cite{solovyev_prb_86_2012magnetic} using first-principles methods that the different magnetic ground state models for  $h$YMnO$_3$ and $h$LuMnO$_3$ have close-by energies. In doped compounds of $h$YMnO$_3$ and $h$LuMnO$_3$, refinement using combination of different representations have been used to get a better result\cite{park_prb_82_2010doping,namdeo_jap_116_2014disorder}.
\\
%
% #######################################################################################################
\begin{figure}[!t]
\centering
\includegraphics[scale=0.40]{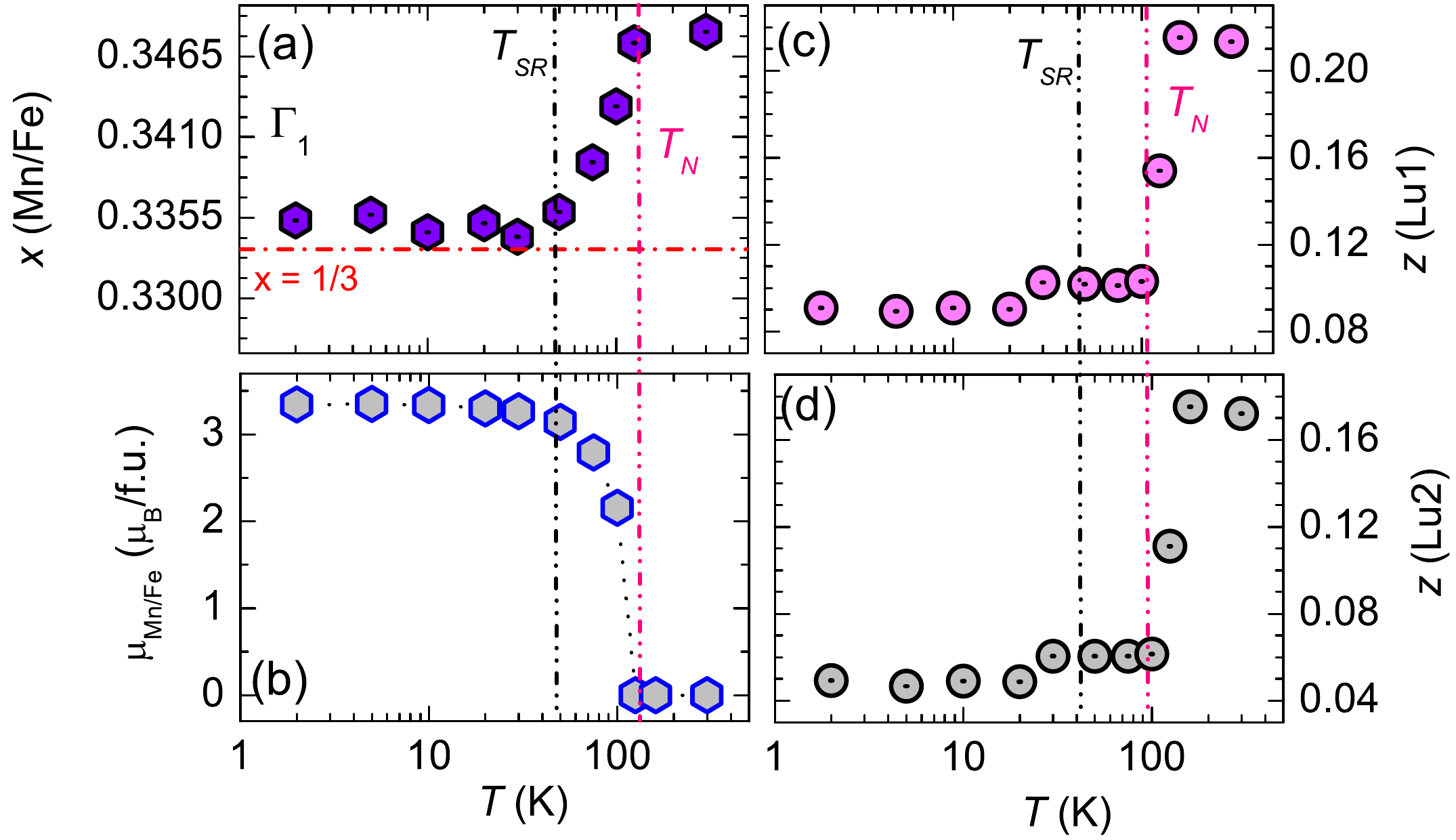}
\caption{(colour online) The refined atomic positions, $x$ for Mn/Fe and $z$ for Lu1 and Lu2 are shown in (a), (b) and (c) respectively. The horizontal line in (a) corresponds to the ideal value of $x$ = $\frac{1}{3}$ for ideal 2D triangular network. All the three show an anomalous change at $T_{\mathrm {SR}}$ indicating strong spin-lattice coupling. The $T_{SR}$ and the $T_N$ are marked as black dash-dotted lines. (d) Shows the variation of ordered magnetic magnetic moment as a function of temperature. Error bars were typically of the size of the data point.}
\label{fig_atoms}
\end{figure}
% #######################################################################################################
\indent 
In Fig~\ref{fig_atoms} (a) (b) and (c), the refined atomic positions, $x$ of Mn/Fe and $z$ of Lu1 and Lu2 are plotted as a function of temperature. The atomic displacements are significantly large and comparable to the giant atomic displacements found in $h$YMnO$_3$\cite{leenature_451_805_2008}. Especially, the $x$ position deviates from the ideal value of $\frac{1}{3}$ for ideal 2D triangular network. Note that (a) presents an anomaly at $T_{\mathrm{SR}}$ and suggest strong spin-lattice coupling. Huge atomic displacements of all the atoms in the unit cell were found to occur below $T_{\mathrm N}$ in multiferroic $h$(Y/Lu)MnO$_3$ leading to strong spin-lattice coupling\cite{leenature_451_805_2008}. The relative shift in Mn $x$ amounts to about 4~$\%$ which is comparable to the values found for $h$YMnO$_3$\cite{leenature_451_805_2008} or for Ti displacements in a conventional ferroelectric like BaTiO$_3$\cite{wyckoff_book1986}. It is interesting to note that the variation of Mn $x$ in YMnO$_3$ and LuMnO$_3$ are in opposite directions, meaning, in one case the $x$ value increases from the ideal $x$ value where as in the other, it decreases. In the case of LuFe$_{0.5}$Mn$_{0.5}$O$_3$, the Mn $x$ variation resembles closely that of YMnO$_3$\cite{leenature_451_805_2008}. In (d), the temperature-evolution of magnetic moment is presented which shows a continuous reduction in magnetic moment and confirms a phase transition at $T_N$, however, at $T_{SR}$ no anomaly is present. The ordered magnetic moment in LuFe$_{0.5}$Mn$_{0.5}$O$_3$ at 2~K estimated from the neutron diffraction data is 3.3~$\mu_B$/f.u. which is comparable to the theoretical value of 3~$\mu_B$ for $h$LuMnO$_3$\cite{parkprb_68_104426_2003magnetic}.
 \\
 % #######################################################################################################
 \begin{figure}[!b]
 \centering
\includegraphics[scale=0.35]{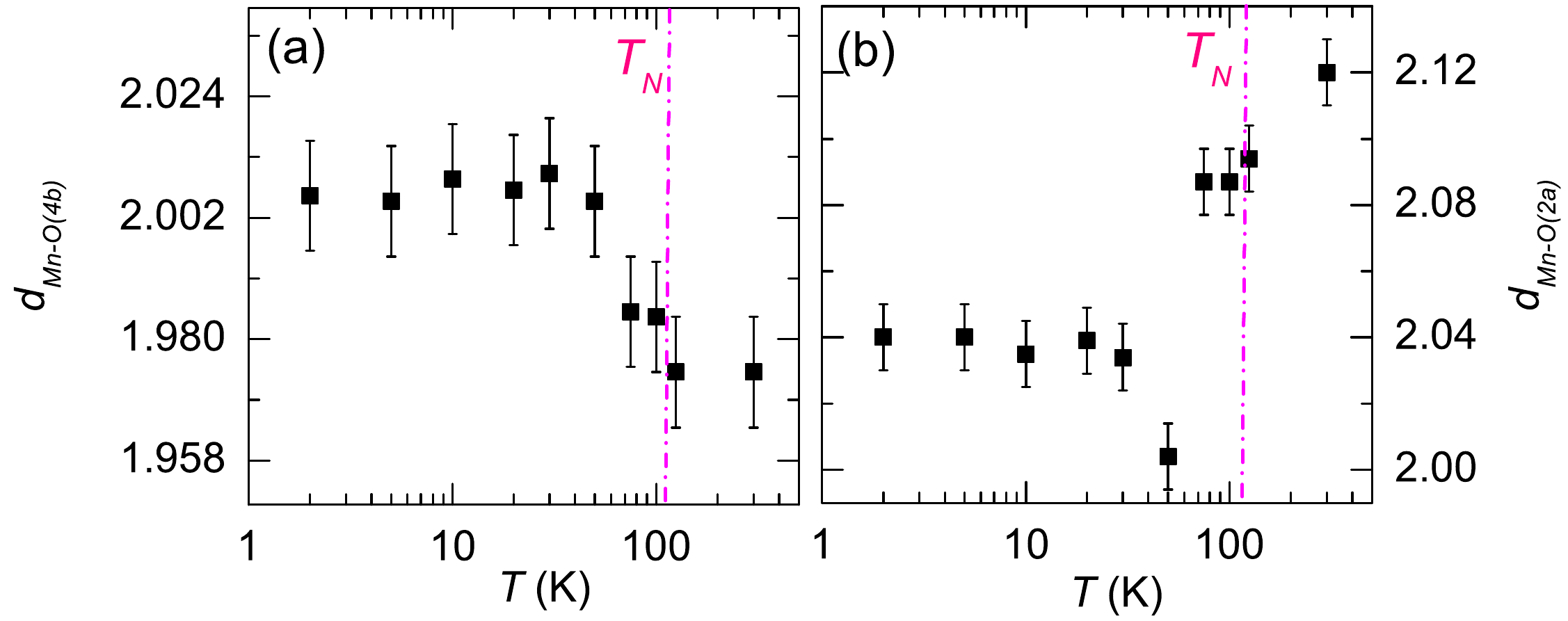}
 \caption{(colour online) The refined bond lengths (Mn/Fe)-O($4b$) and (Mn/Fe)-O($2a$) for LuFe$_{0.5}$Mn$_{0.5}$O$_3$ are presented in (a) and (b) respectively. The anomalies present in Fig~\ref{fig_atoms} are reflected in the bond distances as well, especially close to $T_N$, marked by a vertical dotted line.}
 \label{fig_bonds}
 \end{figure}
 % #######################################################################################################
 %
 %
% #######################################################################################################
\begin{figure}[!t]
\centering
\includegraphics[scale=0.045]{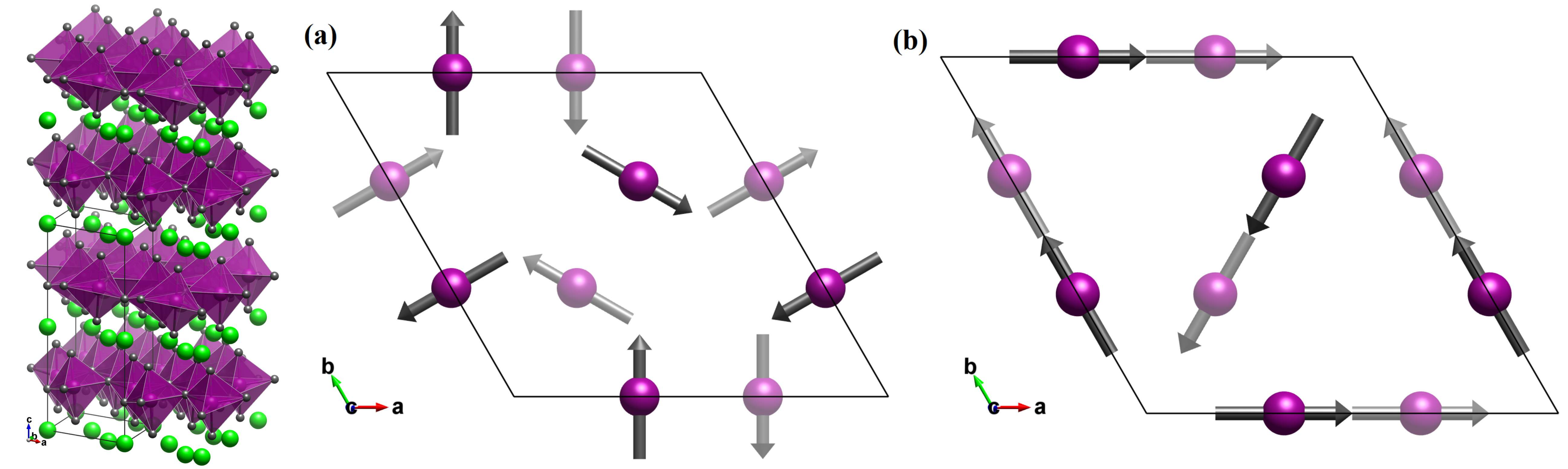}
\caption{(colour online)\label{fig_magneticstr} (Left:) A schematic of the hexagonal crystal structure of LuFe$_{0.5}$Mn$_{0.5}$O$_3$ at room temperature in the $P6_3cm$ symmetry. The Lu atoms are shown in green, the oxygen atoms in grey and the MnO$_5$ polyhedra in violet. The unit cell is outlined using a solid line. (Middle, Right:) The possible magnetic arrangements of Mn spins in LuFe$_{0.5}$Mn$_{0.5}$O$_3$ {\em viz.,} (a) $\Gamma_1$ ($P6_3cm$) and (b) $\Gamma_3$ ($P6'_3cm'$). Only $\Gamma_1$ ($P6_3cm$) structure permits magneto-electric coupling. Both the figures were prepared using VESTA\cite{vesta3}.}
\end{figure}
% #######################################################################################################
%
%
% DISCUSSION
\indent
From the magnetization data on LuFe$_{0.5}$Mn$_{0.5}$O$_3$, it is clear that two magnetic transitions take place in this hexagonal manganite, at $\approx$112~K and $\approx$55~K. The transition at 112~K is confirmed as a paramagnetic-to-antiferromagnetic phase transition. The neutron diffraction data confirms the room temperature nuclear structure as $P6_3cm$. Further, the 2~K data is analyzed faithfully using two magnetic representations - the $\Gamma_1$ ($P6_3cm$) and $\Gamma_3$ ($P6'_3cm'$). The presence of strong intensity for $\bf(100)$ reflection rules out the $\Gamma_2$ ($P6_3c'm'$) and $\Gamma_4$ ($P6'_3c'm$) models. Both the models $\Gamma_1$ and $\Gamma_3$ gave reasonable and comparable reliability factors of refinement. However, a reasonable value for the Mn/Fe $x$ position was obtained only with the $\Gamma_1$ model. For the other model, the refined $x$-value was largely off from the ideal value of $\frac{1}{3}$. In addition, magneto-electric coupling in hexagonal manganites is only allowed for magnetic structures where the six-fold symmetry axis is not combined with time reversal symmetry\cite{brownjpcm_18_10085_2006neutron}. This excludes $\Gamma_3$ ($P6_3'cm'$), $\Gamma_4$ ($P6_3'c'm$) and $\Gamma_6$ ($P6_3'$) structures\cite{brownjpcm_18_10085_2006neutron}. Hence it can be concluded from the observation of significant atomic displacements in the magnetic phase, that the magnetic structure of LuFe$_{0.5}$Mn$_{0.5}$O$_3$ must be $\Gamma_1$ below 112~K. In a recent report, Disseler {\em et al.,} studied the magnetic structure of LuFe$_{0.75}$Mn$_{0.25}$O$_3$\cite{disseler_arxiv_2014magstructure}. They observed a $T_N \approx$ 134~K and the presence of scattering intensity at $\bf(100)$ and $\bf(101)$ reflections above the $T_N$ suggesting that correlations related to both $\Gamma_1$ and $\Gamma_2$ were present. First-principles calculations on hexagonal ferrites and manganites\cite{das_nature_5_2014bulk} do indicate that the energy of $\Gamma_1$ and $\Gamma_2$ are close and are separated by an amount smaller that the single ion anisotropy.
\\
Though giant displacements of the atomic positions of Mn and Lu are observed at $T_{SR}$, such an anomaly is not reflected in the temperature variation of magnetic moment. Hence, it is not possible to confirm a possible change of magnetic structure between $\Gamma_1$ and $\Gamma_3$ at or below 55~K. Detailed studies employing single crystals of this composition is required to settle that question. The important result of our work is the observation of giant atomic displacements across the magnetic transition thereby suggesting strong spin-lattice coupling. In order to confirm the effects of atomic displacements that is observed in LuFe$_{0.5}$Mn$_{0.5}$O$_3$ as presented in Fig~\ref{fig_atoms}, the bond distances (Mn/Fe)-O were evaluated from Rietveld refinement results. The Mn/Fe-O($4b$) and Mn-O($2a$) bond distances which are in the basal plane of the hexagonal structure and presented in Fig~\ref{fig_bonds} (a) and (b) respectively. It is hence clear that the atomic displacements also reflect in the bond distances and is inherent. A schematic of the crystal structure of LuFe$_{0.5}$Mn$_{0.5}$O$_3$ in $P6_3cm$ space group is presented in Fig~\ref{fig_magneticstr} along with $\Gamma_1$ and $\Gamma_3$ structures.
\\
% CONCLUSIONS
\indent 
The magnetic properties of the hexagonal manganite LuFe$_{0.5}$Mn$_{0.5}$O$_3$ are studied in this paper using magnetic measurements and neutron powder diffraction. Frustration effects and, importantly, strong spin-lattice coupling are revealed as a result. It is found that the magnetic structure changes from $\Gamma_4$ for $h$LuMnO$_3$ to $\Gamma_1$ representation for LuFe$_{0.5}$Mn$_{0.5}$O$_3$ while the nuclear structure remains in $P6_3cm$ space group. By perturbing the Mn lattice in the $ab$ plane through substitution of Fe, the temperature of magnetic ordering is enhanced. The atomic positions undergo significant displacements, especially close to $T_N$ and $T_{SR}$ thereby suggesting strong spin-lattice coupling. Our work attains significance following the recent discovery of room temperature multiferroicity in thin films of $h$LuFeO$_3$.
\\ \\
% ACKNOWLEDGEMENTS
AMS thanks the SA-NRF (93549) and the FRC/URC of UJ for financial assistance. HSN acknowledges FRC/URC for a postdoctoral fellowship. HSN wishes to thank Yixi Su for suggesting this compound.
%
%
%
% REFERENCES
%
%
%\bibliography{LFMO}
%\bibliographystyle{apsrev}
%

%
\end{document}